\documentclass[journal,10pt]{IEEEtran}

\usepackage{amsmath,amssymb,amsfonts}
\usepackage{graphicx}
\usepackage[caption=false,font=footnotesize]{subfig}
\usepackage{booktabs}
\usepackage{multirow}
\usepackage{placeins}
\usepackage{xcolor}
\usepackage[colorlinks=true,linkcolor=blue,citecolor=blue,urlcolor=blue]{hyperref}

\graphicspath{{./}}
\newcommand{\crest}{\textup{\textsc{CREST}}}


\begin{document}

\title{CREST: A Hardware-Aware GNN-Based Framework for High-Fidelity and Robust Superconducting Quantum Compilation}

\author{Wenjie~Sun, Xiaoyu~Li, Zhigang~Wang, Yuchen~He, Shengbin~Wang, and~Hantao~Sun
\thanks{Corresponding author: Xiaoyu Li.}%
\thanks{Wenjie Sun and Zhigang Wang are with the School of Electronic Science and Engineering, University of Electronic Science and Technology of China, Chengdu, Sichuan, China (e-mail: \mbox{202211022528@std.uestc.edu.cn}; \mbox{zhigangwang@uestc.edu.cn}).}%
\thanks{Xiaoyu Li is with the School of Information and Software Engineering, University of Electronic Science and Technology of China, Chengdu, Sichuan, China (e-mail: \mbox{xiaoyuuestc@uestc.edu.cn}).}%
\thanks{Yuchen He, Shengbin Wang, and Hantao Sun are with the Quantum Research Institute of China Telecom, China Telecom Quantum Information Technology Group Co., Ltd., Hefei 230088, Anhui, China (e-mail: \mbox{heyuchen@chinatelecom.cn}; \mbox{wangshengbin@chinatelecom.cn}; \mbox{sunhantao@chinatelecom.cn}).}}

\maketitle

\begin{abstract}
	Quantum compilation on superconducting processors requires selecting a connected execution region whose topology and calibrated properties suit the input circuit. This choice constrains subsequent routing and influences the noise encountered during execution, yet aggregate hardware-quality metrics alone may not reliably predict the execution fidelity of a given circuit. Calibration updates can also change region suitability and trigger the rebuilding of calibration-dependent resource structures. We present CREST, a graph neural network (GNN)-based framework that evaluates circuit–region compatibility before routing. It encodes gate dependencies and logical-qubit interactions alongside each candidate region’s connectivity and current calibration attributes; bidirectional cross-graph attention relates the two representations, while real-device execution outcomes supervise candidate ranking. The trained scorer is reused with updated calibration attributes. Evaluations on 324 circuits from seven datasets on TianYan-176 show that CREST improves mean measurement-distribution fidelity by 19.03\% over the strongest baseline and achieves the highest mean fidelity on every dataset. Furthermore, using updated calibration attributes, the same trained model maintains the highest overall mean fidelity across six measurement dates without retraining. With a unified model of circuit–region compatibility, CREST improves execution fidelity across diverse workloads and sustains that advantage as hardware calibration changes.
\end{abstract}


\section{Introduction}
\label{sec:intro}

\IEEEPARstart{Q}{uantum} computing offers opportunities in simulation, optimization, and other computationally demanding applications. Progress on superconducting processors has made the quality of hardware execution an increasingly important concern \cite{arute2019quantum,harrigan2021quantum,kjaergaard2020superconducting}. Gate errors, limited coherence, and imperfect measurement constrain the results obtainable from these devices~\cite{preskill2018quantum}. Quantum compilation is therefore central to their practical use: it translates a logical program into native operations and assigns those operations to physical resources. The compiler's choices determine how much additional work is needed to satisfy hardware constraints and how much noise the program encounters during execution.

Execution-region selection identifies a connected subset of physical qubits within which subsequent placement and routing are performed. A superconducting chip has sparse coupling, so interactions between nonadjacent qubits require extra routing operations. Its physical qubits and couplers also differ in gate error, coherence, and measurement quality~\cite{tannu_not_2019}. Choosing a connected subset of the chip determines both the resources available to a router and the physical errors encountered by the resulting circuit. A poor choice can sacrifice execution fidelity even when the subsequent router performs well within that region.

Prior work has improved quantum compilation through advances in qubit placement, routing, and hardware resource selection. Mapping and routing heuristics reduce added operations through search over placements and upcoming interactions \cite{li2019tackling,zulehner2018efficient,cowtan2019qubit,huang2024routing,fu_effective_2025}, while layout synthesis optimizes explicit gate and depth objectives~\cite{tan2020olsq}. Noise-adaptive, crosstalk-aware, and coherence-aware methods incorporate physical error information into mapping \cite{murali2019noise,wagner2024optimized,tram2025}. At the resource-selection level, QSteed organizes calibrated subregions as virtual processors~\cite{xu2025qsteed}, and HAQA identifies hardware communities using connectivity and fidelity~\cite{sun2026haqa}. Learning-based approaches also guide mapping decisions and reduce search effort \cite{saravanan_noise_2024,sun_mlqm_2025}. Together, these methods demonstrate the importance of both hardware quality and circuit placement. Two practical challenges remain in selecting suitable execution regions for specific circuits.

\emph{First, selecting an execution region solely by aggregate hardware-quality metrics or minimizing routing overhead alone does not guarantee high execution fidelity for a given circuit.} The suitability of a region depends on how its connectivity and local error characteristics interact with the operation sequence and repeated qubit interactions of the circuit. A region with favorable aggregate calibration metrics may still require substantial routing, whereas a routing solution with fewer added operations may repeatedly use less reliable qubits or couplers. Effective region selection therefore requires evaluating circuit structure and calibrated hardware properties together.

\begin{figure}[!t]
\centering
\includegraphics[width=0.94\linewidth]{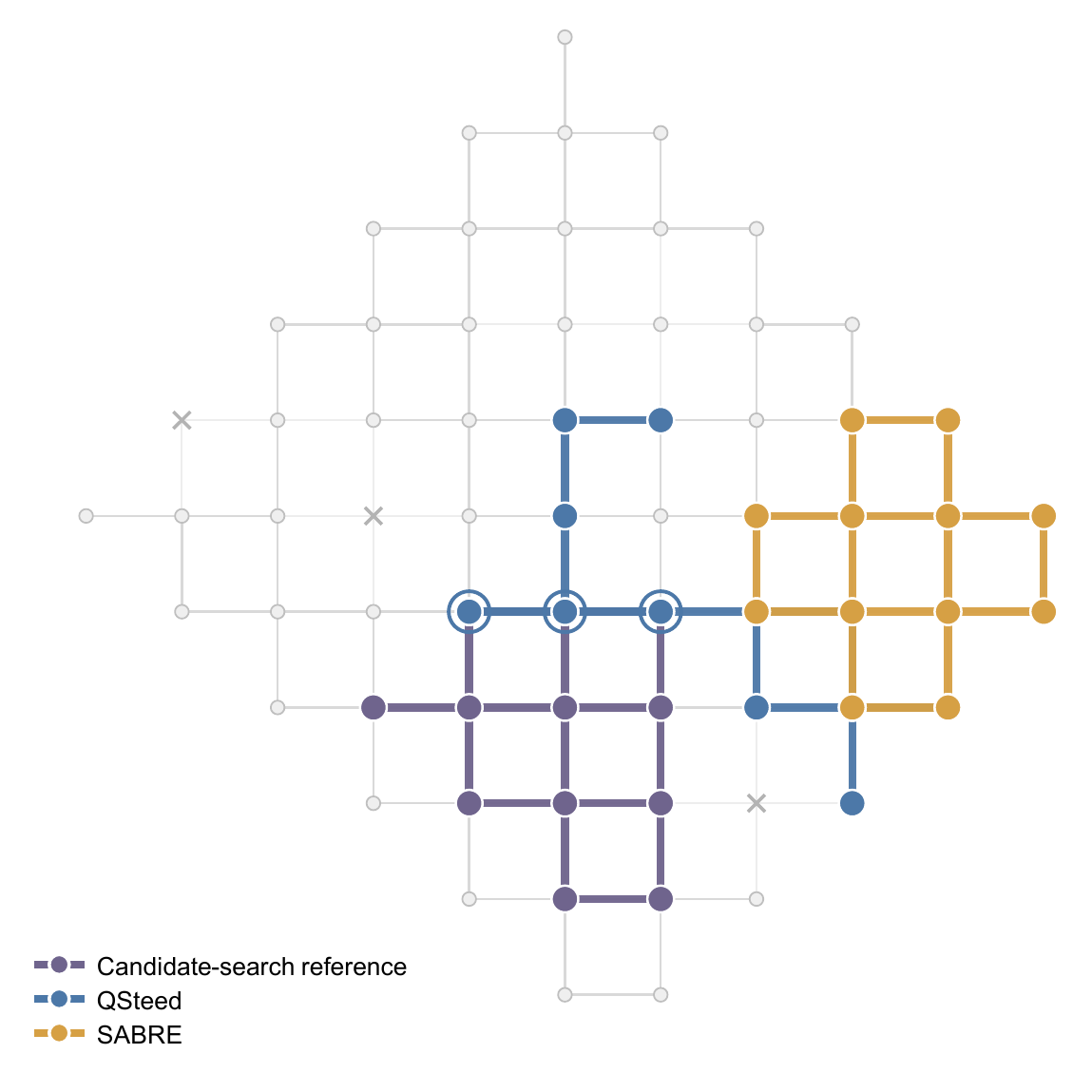}
\caption{Region choice changes the execution quality of the same 12-qubit three-regular circuit on TianYan-176. QSteed selects the blue region, full-graph SABRE uses the gold footprint, and a finite candidate search identifies the violet reference. A local noise model derived from the same calibration gives measurement-distribution fidelities of 0.418, 0.407, and 0.465, with 66, 36, and 42 native CZ gates, respectively.}
\label{fig:motivation}
\end{figure}

Fig.~\ref{fig:motivation} makes this consequence concrete. For the same circuit, a hardware-quality-oriented choice and a full-graph routing result attain simulated fidelities of 0.418 and 0.407. Searching 180 connected, 12-qubit regions yields a reference with fidelity 0.465. That reference uses 42 native CZ gates, compared with 66 and 36 for the other two choices. Thus, even the placement with the fewest entangling gates leaves a substantial fidelity opportunity unused. The example motivates evaluating the compatibility of a circuit and a region together, with execution quality as the target of that evaluation.

\emph{Second, region-selection methods must remain effective across calibration updates while limiting repeated preparation.} Changes in calibrated gate errors, coherence times, and readout errors can alter the suitability of candidate regions even when the coupling topology remains unchanged. Methods that precompute calibration-dependent resource hierarchies may require rebuilding these structures after updates, adding preparation overhead before subsequent circuits can be served. Maintaining effective region selection with current calibration data and limited additional preparation is therefore an important deployment requirement.

To address these problems, we present \crest{}, a hardware-aware, GNN-based framework for superconducting quantum compilation. It integrates multiscale candidate construction, circuit-conditioned region scoring, and compilation within the selected region. The scoring network encodes the original circuit's gate dependencies and logical interactions together with each region's topology and calibrated physical properties. Bidirectional cross-graph attention relates circuit demand to available resources, while real-device outcomes supervise candidate ranking toward higher execution fidelity.

For continued deployment, \crest{} represents current hardware conditions through its candidate representations while sharing the learned circuit--hardware interaction model across calibration snapshots. Valid physical regions can be retained with refreshed attributes, and the same trained scorer evaluates them for incoming circuits without retraining. This design enables reuse of the learned selection capability as hardware conditions evolve. We evaluate this reuse across calibration dates to determine whether the framework retains its execution-fidelity advantage.

This paper makes three contributions:
\begin{itemize}
	\item \emph{A hardware-aware, GNN-based compilation framework.} We integrate multiscale candidate construction, learned circuit--region scoring, and in-region compilation. Hierarchical circuit representations and calibrated region graphs are coupled through bidirectional cross-graph attention, while real-device outcomes supervise region selection toward higher execution fidelity.
	
	\item \emph{A reusable scoring design across calibration updates.} We separate the learned circuit--hardware interaction model from the hardware state supplied through candidate representations. The same trained evaluator can assess current circuit--region pairs without retraining, supporting continued deployment as calibration changes.
	
	\item \emph{Real-device evidence of fidelity and temporal robustness.} Evaluations on 324 circuits from seven datasets on TianYan-176 show a 19.03\% improvement in mean measurement-distribution fidelity over the strongest baseline, with the highest mean fidelity on all seven datasets. Without retraining, the same model maintains the highest overall mean fidelity across six measurement dates.
\end{itemize}

\section{Background and Related Work}
\label{sec:background}

\subsection{Quantum Compilation on Superconducting Hardware}

Quantum programs manipulate qubits through sequences of gates and extract classical results by measurement. A logical circuit describes these operations independently of their physical implementation. Executing it on a superconducting processor requires quantum compilation: gates must be expressed in the processor's supported basis, logical qubits assigned to physical qubits, and interactions made compatible with the available couplers~\cite{sivarajah2021tket}. These transformations preserve the computation while determining the physical operations that accumulate noise during execution.

A processor's coupling graph $G_h=(V_h,E_h)$ records its physical qubits and supported two-qubit interactions. An initial layout associates logical qubits with vertices. If a required interaction spans nonadjacent vertices, routing moves logical states through adjacent SWAP operations and updates their physical locations~\cite{siraichi2018allocation}. Gate synthesis then expresses the resulting operations in the native basis. For example, a SWAP can be decomposed into three CX gates; on a CZ-native processor, a CX is implemented by a CZ bracketed by Hadamard operations on its target, with the single-qubit operations further expressed in the supported rotation basis. Calibration provides gate errors, durations, coherence times, and readout errors for the physical resources used by these operations~\cite{krantz2019guide}.

\begin{figure}[!t]
\centering
\subfloat[]{\includegraphics[width=0.2457\linewidth]{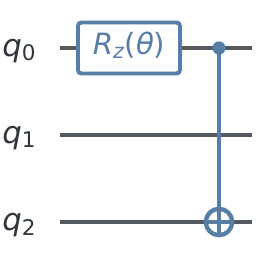}\label{fig:compilation_input}}\hfill
\subfloat[]{\includegraphics[width=0.1429\linewidth]{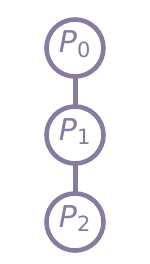}\label{fig:compilation_hardware}}\hfill
\subfloat[]{\includegraphics[width=0.5943\linewidth]{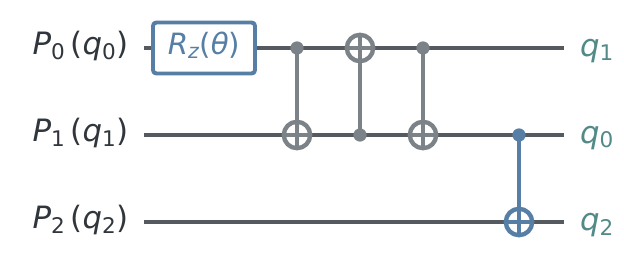}\label{fig:compilation_output}}
\caption{A three-qubit compilation example: (a) logical circuit, (b) coupling graph, and (c) compiled circuit in the CX basis. Parentheses at the wire inputs identify the initial logical qubits assigned to each physical qubit; output labels show the final layout. The first three CX gates in (c) implement a SWAP on $(P_0,P_1)$, enabling the final CX on $(P_1,P_2)$.}
\label{fig:compilation_example}
\end{figure}

Fig.~\ref{fig:compilation_example} illustrates these steps for a circuit containing $R_z(\theta)$ on $q_0$ followed by $\mathrm{CX}(q_0,q_2)$. Under the initial layout $q_i\mapsto P_i$, a linear processor connects $P_0$--$P_1$--$P_2$ and cannot execute the CX directly. A SWAP on $(P_0,P_1)$ moves $q_0$ beside $q_2$; the CX then acts on $(P_1,P_2)$. The compiled circuit contains four nearest-neighbor CX gates, including the three used for the SWAP. Its final layout records the exchanged logical states, allowing measurements to be returned in the original logical order. This small example exposes the compiler's practical tradeoff: satisfying sparse connectivity adds operations, and their physical locations determine the noise encountered by the program.

\subsection{Related Work}

Research on quantum compilation has progressively expanded the information used to make physical placement decisions. Early architecture-aware mapping focused on realizing logical interactions under sparse connectivity while limiting the transformations introduced by routing. Zulehner \emph{et al.} use lookahead to guide mapping to restricted architectures~\cite{zulehner2018efficient}. SABRE combines local SWAP search with forward and backward circuit traversals to improve initial placement~\cite{li2019tackling}, while Cowtan \emph{et al.} develop architecture-agnostic routing informed by upcoming interactions~\cite{cowtan2019qubit}. These approaches make circuit structure central to the search: the value of a placement depends on the interactions that remain to be executed.

Subsequent work improves both the search objectives and the organization of routing. Optimal layout synthesis explicitly optimizes gate and depth costs through mathematical formulations~\cite{tan2020olsq}. More recent routing frameworks dynamically expose executable gates~\cite{huang2024routing}, and parallel mapping coordinates compilation decisions to improve depth and classical efficiency~\cite{fu_effective_2025}. These developments reduce the cost of satisfying hardware constraints. The relation between that cost and physical execution quality, however, also depends on which qubits and couplers carry the resulting operations.

A complementary line therefore incorporates measured hardware properties into compilation~\cite{murali2020crosstalk}. Noise-adaptive mapping uses calibration data to account for differences in gate and measurement reliability~\cite{murali2019noise}. Later approaches model additional physical effects, including crosstalk~\cite{wagner2024optimized} and transverse relaxation~\cite{tram2025}. This progression broadens the compiler's objective from the number of inserted operations to the quality of the operations that will actually execute. It also introduces a changing input: calibration can alter the relative attractiveness of physical resources even when their connectivity is unchanged.

Learning-based methods provide another way to guide this increasingly complex decision process. Saravanan and Saeed combine reinforcement learning and graph neural networks for noise-adaptive mapping~\cite{saravanan_noise_2024}. MLQM uses learned pruning to accelerate optimal qubit mapping~\cite{sun_mlqm_2025}. In these approaches, learned information guides mapping or its search. Region selection offers a related decision at an earlier stage: choosing a connected execution space before detailed placement and routing determines which physical alternatives remain available to those procedures.

Recent resource-oriented frameworks make this earlier decision explicit. QSteed constructs a database of calibrated virtual processors and confines compilation to a selected subregion~\cite{xu2025qsteed}. Its substructure construction includes fidelity-first, degree-first, and randomized growth. Selection can prioritize aggregate two-qubit fidelity or compare the circuit interaction graph with candidate topologies using isomorphism and graph similarity. QSteed thus brings resource organization, circuit structure, and hardware quality into a practical compilation stack. The selection policies express these considerations through predefined fidelity and structural criteria.

HAQA develops a hardware-guided organization of the region search~\cite{sun2026haqa}. Recursive community fusion builds a hierarchy using connectivity and gate fidelity; community expansion adds neighboring resources to provide routing flexibility. This design connects the quality of a hardware region with the size of the search space supplied to a downstream mapper. Its calibration-dependent hierarchy also makes hardware preparation part of deployment: changes to the attributes used in fusion can require this structure to be reconstructed. QSteed maintains updated calibration through its resource database, illustrating a different organization of the same continuing need to serve circuits on a changing processor.

These approaches bring connectivity, calibration, and circuit structure into region selection, but these quantities do not directly specify how well a particular circuit will execute in a region. A region with high aggregate gate fidelity may still require substantial routing for the circuit’s interaction pattern. Conversely, similar interaction graphs can arise from circuits with different gate orders and repeated interactions, which affect their exposure to hardware noise. The remaining problem is therefore to assess circuit–region compatibility at the level of execution outcomes, rather than infer it solely from separately defined hardware and structural criteria.
This assessment must also account for changing calibration. A region’s suitability can change even when its connectivity remains fixed, so a learned selector should incorporate current hardware attributes while retaining the compatibility knowledge acquired from previous executions. 

These considerations motivate \crest{}, which learns to evaluate circuit–region pairs from real-device outcomes. Multiscale candidate construction provides alternative execution spaces, and a bidirectional graph network relates circuit demand to each region’s topology and calibration attributes. The resulting evaluator is reused with updated hardware inputs, without retraining. The following section presents this design.

\section{Methodology}
\label{sec:method}

\subsection{Motivation}
\label{sec:motivation}

Region selection depends on how a circuit's operation sequence uses a candidate region's connectivity and calibrated resources. The key challenge is to assess each candidate in the context of the circuit it would execute, before routing commits to a region, while retaining this evaluation capability as hardware calibration evolves.

We therefore formulate region selection as a learned evaluation of circuit--region pairs. Candidate regions spanning multiple sizes and chip locations expose different tradeoffs in routing space, connectivity, and hardware quality. To evaluate these alternatives, we represent the circuit's gate dependencies and logical-qubit interactions alongside each candidate's calibrated coupling graph. Bidirectional cross-graph attention allows the two representations to exchange information: circuit features capture the relevance of available resources to the circuit's operations, while region features capture how well those resources serve the circuit's demands. Measured execution outcomes supervise the resulting scorer, aligning candidate rankings with the output quality achieved by each circuit--region pairing.

A second design consideration is how to reuse the learned evaluator across calibration updates. Shared graph transformations represent circuit structure and calibrated resources using consistent feature definitions across hardware snapshots. Updated calibration changes the qubit and coupler attributes supplied to scoring, while the learned encoders, attention branches, and readout parameters are retained. The architecture therefore applies a common circuit--hardware compatibility function to changing hardware inputs.

The remainder of this section is organized as follows. Section III-B outlines the overall CREST workflow and its three stages. Section III-C describes multiscale candidate-region construction, including connected-region growth and candidate allocation across region sizes. Section III-D presents the circuit--region compatibility network, detailing circuit and hardware encoding, bidirectional cross-graph attention, and residual score fusion. Section III-E explains how real-device execution outcomes supervise pairwise region ranking and guide model training.

\subsection{Overview}
\label{sec:framework}

\begin{figure*}[!t]
\centering
\includegraphics[width=\textwidth]{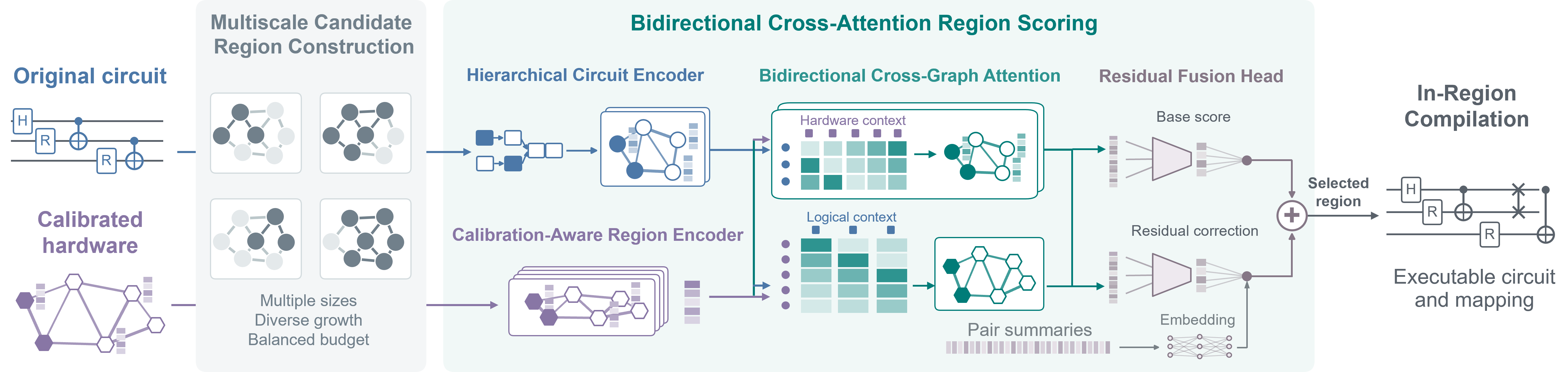}
\caption{Overview of \crest{}. Multiscale Candidate Region Construction provides connected execution spaces with different capacities, locations, and resource characteristics. Bidirectional Cross-Attention Region Scoring evaluates each space in the context of the original circuit, and selects the highest-scoring region. The two attention maps depict opposite query directions; darker cells indicate larger illustrative weights. In-Region Compilation then produces the executable circuit and its physical mapping. The learned scoring stage is expanded in Fig.~\ref{fig:architecture}.}
\label{fig:overview}
\end{figure*}

Fig.~\ref{fig:overview} presents \crest{} as a hardware-aware compilation framework that takes a logical circuit and a calibrated hardware graph and produces an executable program with its physical mapping. \emph{Multiscale Candidate Region Construction} first supplies connected regions with different capacities, locations, and resource characteristics. These alternatives expose different combinations of routing space and hardware quality for the same computation.

\emph{Bidirectional Cross-Attention Region Scoring} provides the framework's GNN-based selection capability. Its Circuit--Region Compatibility Network encodes gate dependencies, logical interactions, and calibrated hardware resources, then relates circuit demand to each candidate through bidirectional cross-graph attention. Trained with real-device outcomes, the network ranks candidates according to their learned compatibility with the circuit and selects the highest-scoring region before detailed placement and routing.

The selected region passes to \emph{In-Region Compilation}, where an existing compiler performs placement, routing, and native-gate synthesis. Calibration updates refresh the hardware attributes used by the same scoring workflow, allowing the framework to reuse its trained selector when evaluating current circuit--region combinations.

\subsection{Multiscale Candidate Region Construction}
\label{sec:candidates}

Candidate construction determines which execution tradeoffs the scorer can consider. A menu dominated by one size or one hardware-quality preference can omit a region well suited to the current circuit. We organize the menu around three complementary choices: region capacity, connected-growth behavior, and the allocation of a bounded candidate budget.

\begin{figure}[!t]
\centering
\includegraphics[width=\linewidth]{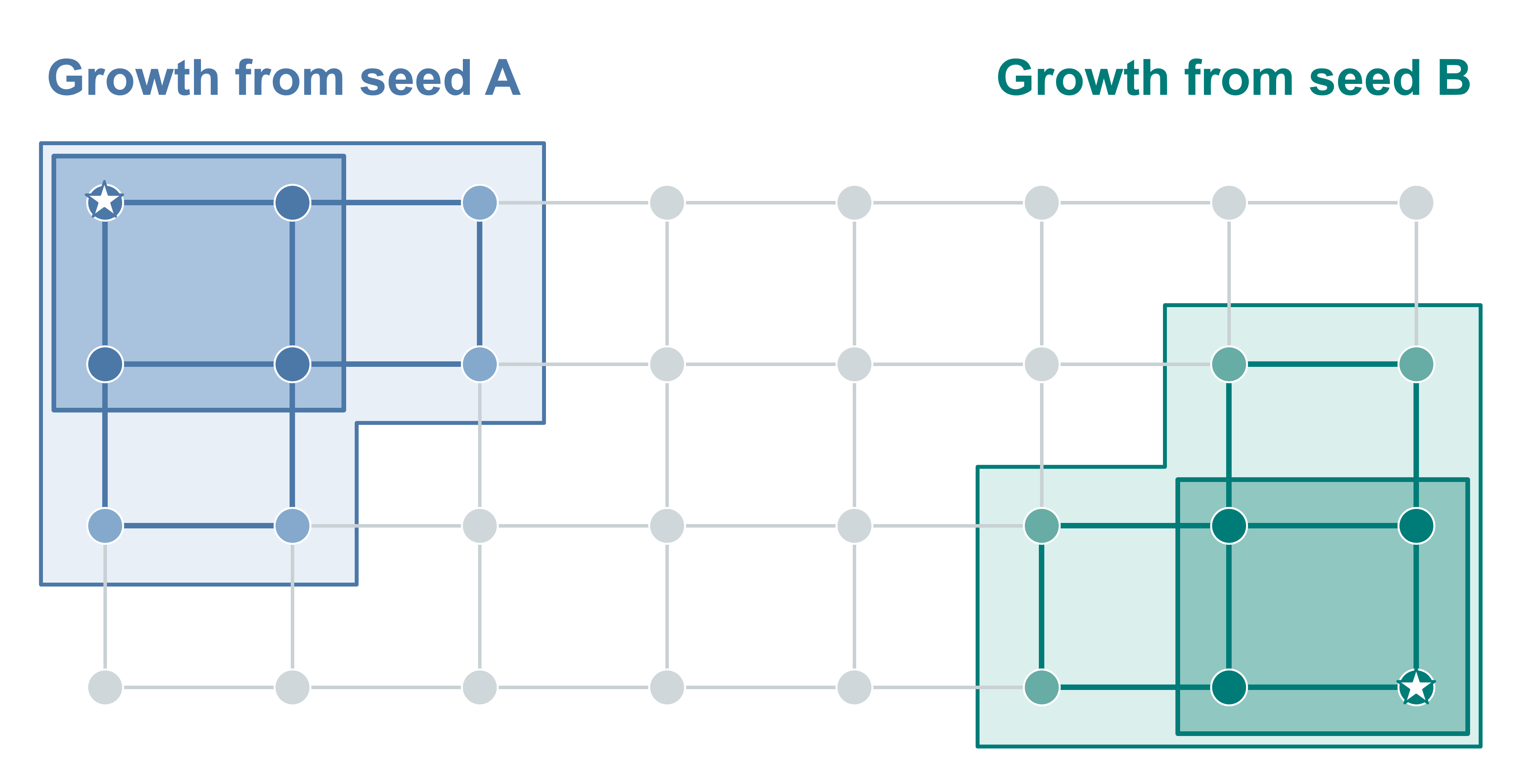}
\caption{Multiscale connected growth at two chip locations. Each color identifies a growth trajectory; its darker compact extent and lighter expanded extent show how additional adjacent resources enlarge the execution space.}
\label{fig:candidate_growth}
\end{figure}

\emph{Multiple capacities for the same circuit.} A width-$n$ circuit can use any connected region with at least $n$ available physical qubits. We include compact regions near this lower bound and larger regions with additional routing space, as illustrated in Fig.~\ref{fig:candidate_growth}. Capacity is therefore a choice presented to the scorer. A larger region opens alternative placements and paths; a compact region concentrates the available resources. Target sizes combine small additive expansions with larger proportional expansions, capped by the available connected component. The evaluated configuration spans the circuit width through a configured upper size of 40 physical qubits.

\emph{Complementary connected-growth policies.} For each target size, growth starts from multiple physical seeds and repeatedly adds a neighboring qubit until the target is reached. Quality-led growth favors low gate and readout error, connectivity-led growth favors internal links and well-connected nodes, and seeded randomized growth explores alternative shapes. Each added node preserves connectivity, and the candidate includes the coupling edges induced by its physical set. Separate runs for different target sizes and policies provide alternative regions; the nested extents in Fig.~\ref{fig:candidate_growth} depict the evolution within illustrative growth runs.

\emph{Balanced allocation across sizes and quality.} We assign a share of the candidate budget to each target size and remove duplicate physical sets. Within a size, sampling alternates between a quality-ranked leading subset and the full pool produced by growth. The leading subset retains favorable hardware options, while the full-pool draws preserve alternatives with different shapes and resource combinations. These draws populate the menu for comparison; the network subsequently chooses one region for execution. Valid previously constructed physical sets can also be retained, and their calibration attributes are refreshed before scoring. In the evaluated configuration, this process provides up to 192 distinct candidates per circuit.

Candidate construction uses the circuit width to determine feasible capacities. Hardware-derived pools can be prepared and reused across circuits of the same width; the original circuit's detailed structure enters the scoring stage. When hardware availability changes, we remove invalid candidates and replenish the menu from the current usable graph. This division supplies diverse feasible choices while leaving their circuit-specific evaluation to the learned network.

\subsection{Bidirectional Cross-Attention Region Scoring}
\label{sec:architecture}
\label{sec:cross_attention}

\begin{figure*}[!t]
\centering
\includegraphics[width=\textwidth]{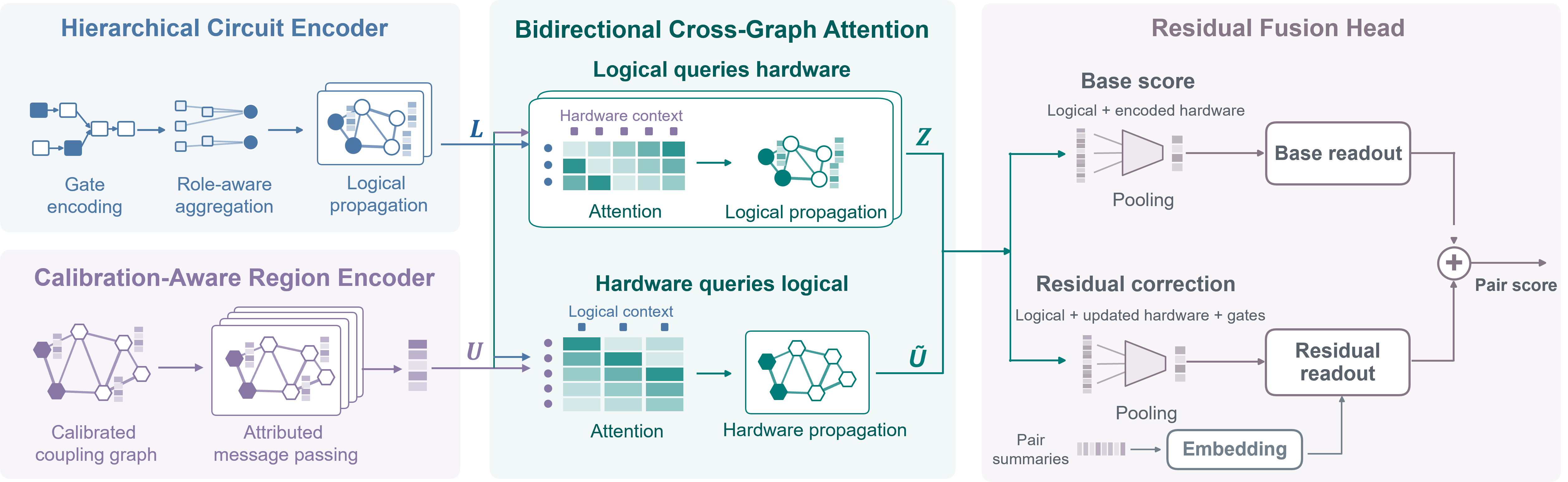}
\caption{Circuit--Region Compatibility Network, expanding the scoring stage in Fig.~\ref{fig:overview} with the same module colors. Directed gate encoding, role-aware aggregation, and logical propagation encode the circuit; calibrated message passing encodes the region. Attention-map rows represent queries and columns represent context nodes. The logical-query branch updates logical features; the hardware-query branch reads the original encoded logical features and updates hardware features. The base readout pools logical and encoded hardware features. The residual readout combines pooled logical, updated hardware, and gate features with a pair-summary embedding. Both readouts use scale descriptors and their outputs are added. Matrix weights and graph stacks are schematic.}
\label{fig:architecture}
\end{figure*}

The Circuit--Region Compatibility Network evaluates how well a candidate's resources support the computation expressed by the original circuit. As shown in Fig.~\ref{fig:architecture}, it comprises a \emph{Hierarchical Circuit Encoder}, a \emph{Calibration-Aware Region Encoder}, \emph{Bidirectional Cross-Graph Attention}, and a \emph{Residual Fusion Head}. The encoders retain the structure of the two inputs; attention relates them before their information is summarized into a selection score.

\subsubsection{Hierarchical Circuit Encoder}
The circuit encoder preserves both the order of operations and the pattern of interactions among logical qubits. Its first level is a directed gate-dependency graph. Gate nodes encode operation type, numerical parameters, arity, and position; dependency edges connect successive operations on a wire and identify operand roles. Directed message passing aggregates predecessor and successor context with separate transformations~\cite{gilmer2017message}. The resulting gate states describe an operation together with the computation around it.

The second level projects these states onto logical qubits through role-aware aggregation. Contributions from a control, target, or other operand are distinguished before aggregation, and logical-node summaries provide operation counts and temporal information. An attributed interaction graph then connects qubits that participate in common gates. Its edges describe interaction frequency, direction, and distribution across circuit phases. GINE message passing~\cite{xu_how_2018,hu2020pretraining} propagates these attributes to produce logical states $L$. This hierarchy connects detailed gate dependencies with a qubit-level representation that can be compared with physical resources.

\subsubsection{Calibration-Aware Region Encoder}
The region encoder represents each candidate as its physical coupling graph. Node attributes describe single-qubit operations, coherence, readout error, and local connectivity. Edge attributes describe native entangling operations, their errors and durations, and missing-value indicators. Shared transformations and physical scales preserve quality differences across candidates. GINE layers aggregate neighboring states with the associated coupling attributes, producing hardware states $U$ that express both local reliability and its connected context.

Hardware identifiers locate the graph and its calibration records; the encoded attributes describe the physical properties used for evaluation. This representation lets the same network process regions with different sizes and apply current calibration through the same input path. A change in a coupler's error or a qubit's coherence thus changes the candidate representation supplied to scoring.

\subsubsection{Bidirectional Cross-Graph Attention}
The two encoded graphs describe potential demand and available resources. Cross-attention~\cite{vaswani2017attention,li2019graphmatching} relates these descriptions before global pooling. Logical states first query the hardware states, allowing each logical qubit to aggregate resource information according to learned compatibility. Each attention update is followed by a residual normalization and propagation over the logical interaction graph. Two successive blocks produce region-conditioned logical states $Z$, sharing resource context among qubits that interact in the circuit.

A complementary branch lets hardware states query $L$, the logical representation produced by the circuit encoder. A residual attention update followed by propagation over the coupling graph produces circuit-conditioned hardware states $\widetilde U$. The two branches answer complementary questions: which resource characteristics matter to the circuit's interacting qubits, and how circuit demand changes the significance of the resources available within the region. Attention supplies continuous representations for evaluating the pair; the downstream compiler constructs the discrete mapping and route.

\subsubsection{Residual Fusion Head}
The scoring head combines these local interactions with global context. A base readout uses mean and maximum pooling of the region-conditioned logical states $Z$ and the encoded hardware states $U$, together with circuit and region scale descriptors. The residual readout adds the circuit-conditioned hardware states $\widetilde U$ and learned attention pooling over logical, hardware, and gate states. Attention pooling allows concentrated interactions and locally significant physical conditions to influence the summary.

The residual path also embeds compact circuit--region statistics, including interaction repetition, regional distances and connectivity, error distributions, coherence, and workload--resource pressure. These summaries complement the graph representations with explicit global properties. A learned correction jointly consumes the pooled graph context, the statistics embedding, and scale descriptors. The final score is
\begin{equation}
s_\theta(C,R)=s_{\mathrm{base}}(C,R)+s_{\mathrm{res}}(C,R).
\label{eq:fusion_score}
\end{equation}
Both terms are trained jointly with the encoders and attention branches. For a circuit $C$ and calibrated hardware graph $G_h$, candidate construction provides the menu $\mathcal{R}(C,G_h)$. After evaluating these candidates, the network selects the highest-scoring region for compilation:
\begin{equation}
	R^*=\arg\max_{R\in\mathcal{R}(C,G_h)}s_\theta(C,R).
	\label{eq:selection}
\end{equation}

The selected region is passed to an existing compilation engine, which performs physical placement, routing, and native-gate synthesis subject to the region’s connectivity and the processor’s native-gate set. This stage returns the executable circuit and the qubit mapping needed to interpret its measurement outcomes.

\subsection{Training with Real-Device Outcomes}
\label{sec:training}

To learn which execution spaces are useful, we train on comparisons between regions that execute the same circuit. Each development circuit is compiled within several candidate regions and run on the processor. Measurement bitstrings are restored to logical order, and the agreement between the measured and ideal output distributions supplies an execution-quality label $H$. This supervision includes the combined influence of the compiled operations and final measurement, linking region evaluation to the quality of the observed result.

Training groups observations by circuit. For two regions $R_i$ and $R_j$, the measured gap $\Delta H_{ij}=H_i-H_j$ defines their preference, and the network produces $\Delta s_{ij}=s_\theta(C,R_i)-s_\theta(C,R_j)$. Comparing alternatives for the same computation focuses learning on the selection decision that deployment will make~\cite{burges2005ranking}. We combine an ordering objective with a robust score-gap objective:
\begin{equation}
\begin{split}
\mathcal L_{ij}=w_{ij}\big[&\operatorname{softplus}\big(-\operatorname{sign}(\Delta H_{ij})\Delta s_{ij}/\tau\big)\\
&+\lambda\,\operatorname{SmoothL1}_{\beta}(\Delta s_{ij},\Delta H_{ij})\big].
\end{split}
\label{eq:pair_loss}
\end{equation}
The first term encourages the correct regional ordering. The second preserves information about the magnitude of the measured quality difference while limiting the influence of large residual errors. The weight $w_{ij}$ increases with the observed quality gap relative to the measurement-uncertainty proxy, reducing the influence of weak or uncertain preferences. The constants $\tau$, $\beta$, and $\lambda$ control the smoothness and balance of the objectives.

We jointly train the gate encoder, graph encoders, attention branches, and fusion head by averaging the loss over candidate pairs and circuit groups. Structurally related circuit variants remain together when forming the training and validation sets. After each training epoch, the model selects one region for each validation circuit. We evaluate these selections using their measured fidelities in the real-device dataset and retain the model with the best validation performance. The development holdout is reserved for final evaluation.

\section{Experimental Evaluation}
\label{sec:experiment}

To evaluate whether \crest{} improves execution fidelity and retains effective region selection as hardware calibration evolves, we design complementary experiments on compilation quality, selection effectiveness, and model reuse. Complete-flow comparisons assess the fidelity benefit of the framework. To identify how region selection contributes to this benefit, we control the routing backend and candidate menu, then use a statistics-only ablation to examine the joint contribution of graph encoding and cross-graph attention. Held-out circuits test transfer beyond the development instances, while repeated evaluation with a fixed model tests robustness across calibration dates. Finally, timing measurements characterize the downstream compilation cost of the regions selected by \crest{}.

\subsection{Experimental Setup}
\label{sec:settings}

\paragraph{Hardware and execution}
To measure the effect of region selection on physical execution, we use the Tianyan quantum-computing platform, which provides cloud access to a family of programmable superconducting systems~\cite{tianyan2025cloud}. Its Tianyan-176 systems use Zuchongzhi-2.0-like processors~\cite{tianyan2025cloud,wu2021strong}. Our experiments use TianYan-176 with 63 available physical qubits and 94 coupling edges. The CQLib SDK~\cite{cqlib2025sdk} provides the connectivity and calibrated gate, readout, and coherence attributes used by the compilation flows. All methods in the primary comparison use the same hardware information. Submitted programs use the QCIS instructions \texttt{RZ}, \texttt{X2P}, \texttt{X}, \texttt{CZ}, and \texttt{M}, corresponding to $R_z(\theta)$, $R_x(\pi/2)$, $X$, CZ, and computational-basis measurement. Each circuit is measured with 1,000 shots, without readout mitigation or postselection.

\paragraph{Implementation} Classical compilation and neural inference run on an Intel Core Ultra X7 358H workstation with 32 GB memory. We use Python 3.10, PyTorch 2.3.1, Qiskit 2.5.2, and QSteed 0.2.2.

\begin{table}[!t]
\centering
\caption{Benchmark Suite}
\label{tab:benchmarks}
\small
\begin{tabular}{@{}lrr@{}}
\toprule
Dataset & Circuits & Logical qubits \\
\midrule
QASMBench & 18 & 4--19 \\
Q-Tetris & 12 & 4--16 \\
MQT Bench & 61 & 5--20 \\
Quantum Volume & 3 & 5--15 \\
Hardware-efficient VQE & 20 & 4--20 \\
Static QEC & 150 & 5--17 \\
Three-regular graphs & 60 & 10--20 \\
\midrule
Total & 324 & 4--20 \\
\bottomrule
\end{tabular}
\end{table}

\paragraph{Benchmarks}
To cover different interaction patterns and program structures, we assemble 324 circuits from seven benchmark collections, summarized in Table~\ref{tab:benchmarks}. QASMBench~\cite{li2023qasmbench} and Q-Tetris~\cite{qtetris} contribute arithmetic, reversible-logic, and other application circuits. MQT Bench~\cite{quetschlich2023mqt} includes Deutsch--Jozsa, GHZ, W-state, and TwoLocal circuits. Quantum Volume~\cite{cross2019validating} exercises randomized interactions, while hardware-efficient VQE repeats parameterized rotations and entangling layers~\cite{kandala2017hardware}. Three-regular instances vary sparse nonlocal interactions. Static QEC circuits contribute coherent preparation and syndrome-related gate structures from repetition, five-qubit, Shor, Steane, and rotated-surface codes~\cite{terhal2015error}. Together, these benchmarks vary both the computation presented to the selector and its demands on physical connectivity. The QEC measurements evaluate compiled terminal distributions rather than logical decoding with real-time feedback.

During the development of \crest{}, 173 of the 324 evaluation circuits were used for training and 55 for validation. The remaining 96 circuits were excluded from both, providing an independent subset for evaluating generalization. We report results on the complete suite and examine this held-out subset separately in Section~\ref{sec:generalization}. The final model is selected using the validation procedure in Section~\ref{sec:training}, then held fixed throughout evaluation.

\paragraph{Baselines}
We compare \crest{} with two state-of-the-art region-selection frameworks, QSteed and HAQA, alongside SABRE as a full-graph routing baseline and Random as a selection control. QSteed~\cite{xu2025qsteed} generates and ranks hardware resources using coupling fidelity, then uses its original compilation backend. HAQA~\cite{sun2026haqa} selects a connected hardware region through its connectivity- and fidelity-aware community procedure, followed by SABRE. SABRE~\cite{li2019tackling} performs layout and routing over the full available graph, providing a reference without explicit region selection. Random selects uniformly from the same valid candidate menu as \crest{} and uses the same SABRE backend, testing the benefit of learned scoring within that menu.

All SABRE-based flows use identical routing settings. QSteed retains its original backend in the complete-flow comparison; Section~\ref{sec:selection_eval} additionally controls the routing backend and candidate menu to distinguish selection quality from other compilation choices. The complete-flow comparisons, selection controls, and ablation all use the same 324-circuit suite.

\begin{figure*}[!tp]
\centering
\includegraphics[width=0.96\textwidth]{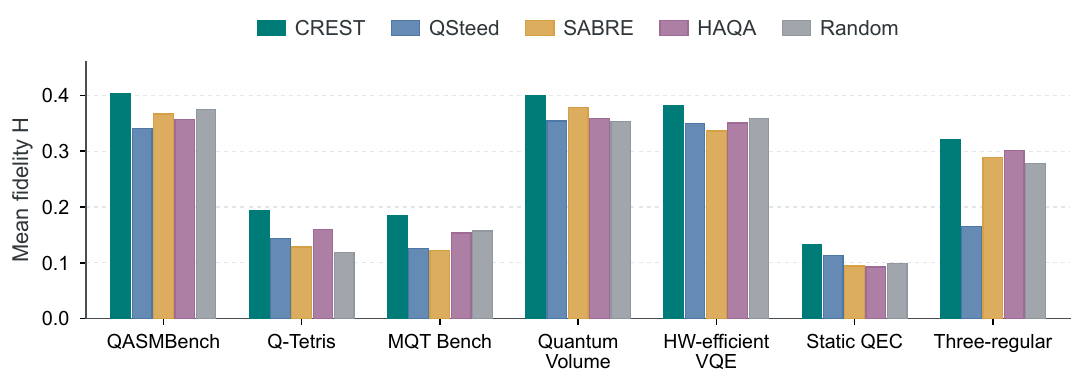}
\caption{Mean measurement-distribution fidelity of the five compilation flows on the seven datasets. Every dataset uses the common set of circuits listed in Table~\ref{tab:benchmarks}, with equal weight per circuit. All bars show fidelity values on TianYan-176.}
\label{fig:datasets}
\end{figure*}

\paragraph{Metrics and aggregation}
Let $p(x)$ be the ideal terminal measurement distribution of the input circuit and $q(x)$ the empirical distribution obtained on the processor. We use measurement-distribution fidelity,
\begin{equation}
H(p,q)=\left(\sum_x\sqrt{p(x)q(x)}\right)^2,
\label{eq:measurement_fidelity}
\end{equation}
which ranges from zero to one and reaches one for matching distributions. This metric captures the combined effect of compiled operations, physical placement, and readout on the circuit's observed output. Ideal distributions are computed from the original logical circuits.

We report mean fidelity with equal weight per circuit and relative improvements over the baseline mean. To estimate how much the mean fidelity difference varies with the sampled circuits, we repeatedly sample circuit groups with replacement and recompute the difference~\cite{efron_bootstrap_1992,field_bootstrapping_2007}. Circuits with the same logical width and ordered gate-name/qubit trace form a group, so structurally related variants are sampled together. Every sample retains all methods' results on the same circuits, preserving the paired comparison. We report the central 95\% interval from 2,000 such bootstrap resamples.

\subsection{Overall Compilation Quality}
\label{sec:main}

\begin{table}[!t]
\centering
\caption{Overall Fidelity on 324 Circuits. $\Delta H$ is the paired mean of CREST minus the indicated method.}
\label{tab:overall}
\small
\begin{tabular}{@{}lrrc@{}}
\toprule
Method & Mean $H$ & $\Delta H$ & 95\% interval \\
\midrule
CREST & \textbf{0.21201} & -- & -- \\
QSteed & 0.15545 & 0.05656 & [0.04588, 0.06787] \\
SABRE & 0.16924 & 0.04277 & [0.03207, 0.05440] \\
HAQA & 0.17811 & 0.03390 & [0.02244, 0.04557] \\
Random & 0.17734 & 0.03467 & [0.02311, 0.04634] \\
\bottomrule
\end{tabular}
\end{table}

Table~\ref{tab:overall} reports the complete-flow results. \crest{} achieves a mean fidelity of 0.21201, followed by HAQA at 0.17811, Random at 0.17734, SABRE at 0.16924, and QSteed at 0.15545. The relative improvement over the strongest overall baseline, HAQA, is 19.03\%. The corresponding improvements over Random, SABRE, and QSteed are 19.55\%, 25.27\%, and 36.38\%, respectively. All four paired mean differences have positive 95\% bootstrap intervals. These results establish a consistent aggregate advantage over both calibration-based region selection and full-graph routing.

Fig.~\ref{fig:datasets} separates the results by dataset. \crest{} has the highest mean fidelity in all seven groups. On QASMBench, it achieves 0.40284, compared with 0.37438 for the strongest baseline, Random. On Q-Tetris, the corresponding values are 0.19361 and 0.15931 for HAQA. MQT Bench reaches 0.18443 with \crest{} and 0.15695 with Random. The relative improvements in these three heterogeneous benchmark collections are 7.60\%, 21.53\%, and 17.51\%. Their circuits differ in gate order and interaction pattern, so the gain spans a diverse application workload.

The advantage also extends to more regular circuit constructions. For Quantum Volume, \crest{} reaches 0.39891 against SABRE's 0.37803, a 5.52\% improvement. Hardware-efficient VQE reaches 0.38095 against Random's 0.35930, improving by 6.03\%. On three-regular graphs, \crest{} achieves 0.32085 while HAQA achieves 0.30155, a 6.40\% improvement. These workloads place different demands on connectivity: randomized circuits require varied interactions, variational circuits repeat an entangling pattern, and three-regular instances require a sparse nonlocal interaction graph. The consistent dataset-level ranking indicates that the selection policy serves multiple interaction regimes on the same processor.

The static QEC group has the lowest aggregate fidelity because it includes long coherent gate sequences. \crest{} achieves 0.13199, exceeding QSteed's 0.11285 by 16.97\%. The aggregate benefit is mainly visible in the repetition-code circuits. Several deeper surface, Shor, and Steane instances produce fidelities close to zero for all methods, where 1,000 shots provide limited resolution. These results describe compilation quality for static QEC gate structures; an evaluation of logical correction would additionally require syndrome decoding and the appropriate execution protocol.

\begin{figure*}[!tp]
\centering
\begin{minipage}{0.485\textwidth}
\centering
\includegraphics[width=\linewidth]{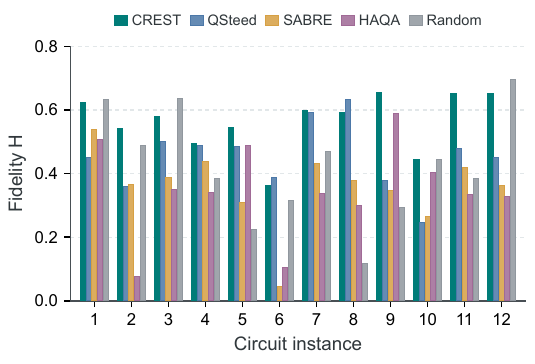}
\par\small (a) X-type checks
\end{minipage}\hfill
\begin{minipage}{0.485\textwidth}
\centering
\includegraphics[width=\linewidth]{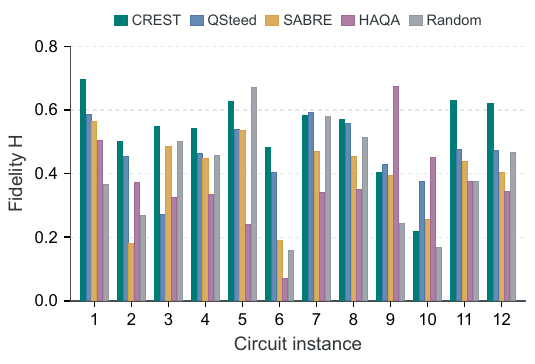}
\par\small (b) Z-type checks
\end{minipage}
\caption{Circuit-level fidelity for three-data-qubit repetition codes with X- and Z-type checks. Each panel contains all 12 instances from three gate schedules and four input states, using five logical qubits. Both panels use the same zero-based vertical scale.}
\label{fig:repetition_instances}
\end{figure*}

\paragraph{Circuit-level behavior}
Fig.~\ref{fig:repetition_instances} examines the repetition-code circuits at a fixed width while varying gate schedule and input state. These variations change the order and location of repeated interactions without changing the number of logical qubits. \crest{} reaches mean fidelities of 0.56155 and 0.53503 for the X- and Z-check groups, compared with 0.45439 and 0.46859 for QSteed, the strongest mean baseline in each group. It exceeds every baseline on 14 of the 24 individual circuits and has the higher mean in both groups. This behavior illustrates how conditioning region choice on the circuit can benefit related computations even when their width is identical; the full-suite results above establish the aggregate comparison across workloads.

\begin{figure*}[!tp]
\centering
\begin{minipage}{0.485\textwidth}
\centering
\includegraphics[width=\linewidth]{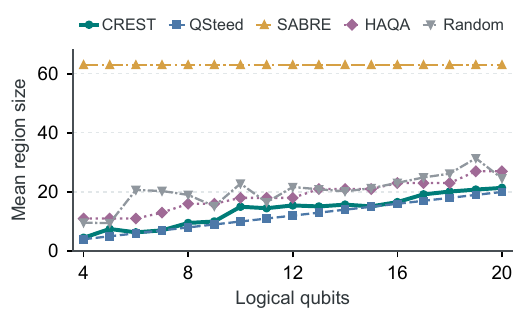}
\par\small (a) Selected region size
\end{minipage}\hfill
\begin{minipage}{0.485\textwidth}
\centering
\includegraphics[width=\linewidth]{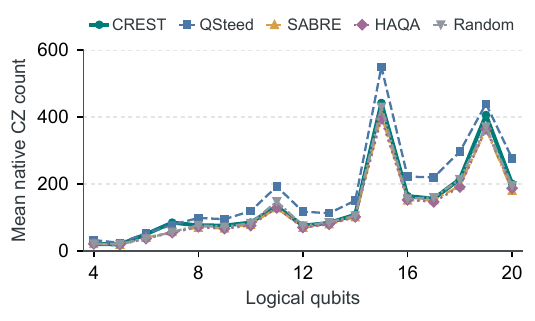}
\par\small (b) Native CZ count
\end{minipage}
\caption{Hardware allocation and compiled two-qubit gate cost across the 324-circuit workload. Each point averages the same circuits at a given logical width for every method. Panel (b) uses a linear vertical axis starting at zero. The workload composition varies across widths.}
\label{fig:resources}
\end{figure*}

To relate the fidelity results to hardware use and compiled gate cost, Fig.~\ref{fig:resources} reports selected region size and native CZ count across all 324 circuits, grouped by logical width. Full-graph SABRE retains all 63 available qubits, while \crest{} allocates a much smaller region that grows with circuit width. Across the suite, \crest{} produces an average of 115.93 native CZ gates, while SABRE and HAQA produce 107.30 and 106.45, respectively. \crest{} nevertheless achieves the highest measured fidelity. Physical region quality therefore adds a material dimension to compilation beyond total two-qubit gate count. At the same time, the mean selected region contains 14.08 qubits, providing a compact routing space within the 63-qubit usable graph. The following controls examine the contribution of choosing that space.

\subsection{Effectiveness of Region Selection}
\label{sec:selection_eval}

Complete compilation flows combine region generation, selection, and routing. We conduct two controls to examine selection at different module boundaries. Both controls keep the \crest{} model fixed and evaluate the same 324 circuits with the execution protocol above.

\begin{figure}[!t]
\centering
\includegraphics[width=0.98\linewidth]{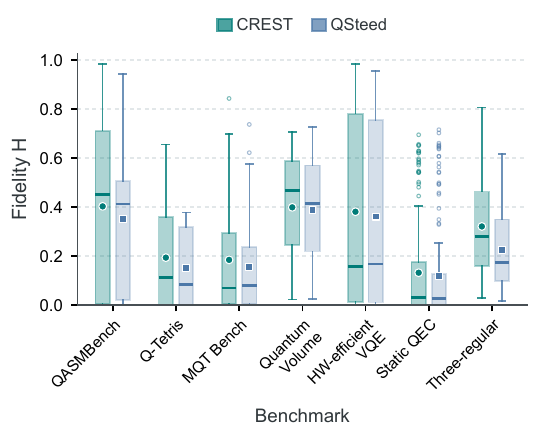}
\caption{Region-selection comparison using a shared SABRE backend. CREST and QSteed retain their respective region-generation and selection procedures; both selected regions are compiled with identical SABRE settings. Boxes show circuit-level fidelity distributions within each benchmark; filled markers show means and hollow points show outliers.}
\label{fig:shared_routing}
\end{figure}

\paragraph{Shared routing backend}
We retain QSteed's original resource-generation and selection stages and route the selected region with the same SABRE configuration used by \crest{}. This setting holds the downstream router constant and compares the region choices made by the two selection pipelines. Fig.~\ref{fig:shared_routing} compares the circuit-level fidelity distributions of the two pipelines by benchmark. \crest{} achieves 0.21201 overall, compared with 0.17699 for QSteed under the shared router. The relative gain is 19.79\%, and the paired difference is 0.03502 with a 95\% interval of [0.02635, 0.04411]. \crest{} has the higher mean in each dataset, including QASMBench at 0.40284 versus 0.35293 and three-regular graphs at 0.32085 versus 0.22415. Thus, the overall advantage persists when both selection pipelines feed the same routing implementation.

\begin{figure}[!tp]
\centering
\includegraphics[width=0.98\linewidth]{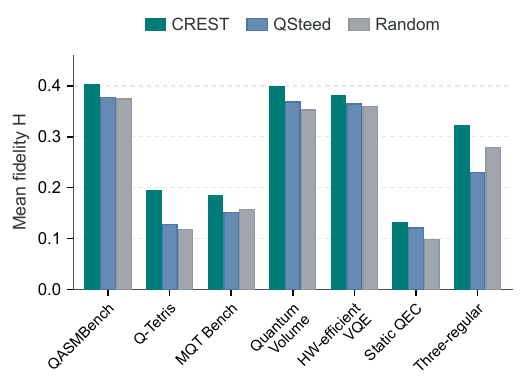}
\caption{Region-scoring comparison with identical candidate menus and SABRE settings. CREST scores circuit--region pairs, QSteed applies its original coupling-fidelity product to each region, and Random samples the menu uniformly. Each method compiles its selected region.}
\label{fig:same_candidates}
\end{figure}

\paragraph{Shared candidate menu}
We next supply \crest{}, QSteed, and Random with the same connected candidate regions for each circuit. QSteed applies its original product of coupling fidelities directly to these regions. All three methods select one candidate and use identical SABRE settings. This control compares the scoring policies while fixing both the available regions and the router.

Fig.~\ref{fig:same_candidates} shows that \crest{} again has the highest mean in every dataset. The overall means are 0.21201 for \crest{}, 0.17855 for QSteed, and 0.17734 for Random. The relative improvement over QSteed is 18.74\%, with a paired difference of 0.03346 and a 95\% interval of [0.02439, 0.04317]. Q-Tetris exhibits a particularly clear separation: \crest{} reaches 0.19361, while QSteed and Random reach 0.12806 and 0.11828. On MQT Bench, the values are 0.18443, 0.15174, and 0.15695. The same pool contains the regions used by every policy, so these differences establish the practical value of the learned selection decision within an available hardware menu.

Taken together, the controls locate a substantial part of the gain at the region-selection stage. A common router preserves the advantage of the \crest{} pipeline, and common candidate menus preserve the advantage of its scorer. The result supports circuit-conditioned evaluation as a useful compilation interface: hardware regions with different connectivity and calibration profiles can be assessed against the gate structure of the current input before routing.

\subsection{Ablation Study}
\label{sec:ablation}

\begin{figure}[!tp]
\centering
\includegraphics[width=0.98\linewidth]{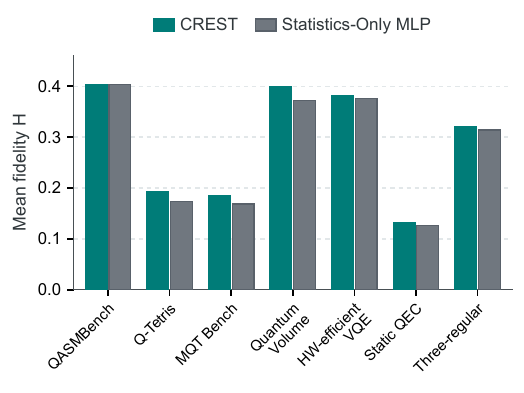}
\caption{CREST and the Statistics-Only MLP on all seven datasets, using identical candidate menus and SABRE settings. The ablation scores the same 28 aggregate circuit--region features without graph encoders or cross-attention. Each mean includes all circuits in the corresponding dataset.}
\label{fig:ablation}
\end{figure}

We compare \crest{} with a \textit{Statistics-Only MLP} to examine what structured representations add beyond aggregate circuit--region features. This independently trained multilayer perceptron uses the same 28 statistics supplied to the fusion head, with the same measured-label training data, circuit splits, and learning objective. Both models score identical candidate menus and compile their chosen region with the same SABRE configuration. The comparison therefore isolates the joint contribution of graph encoding and circuit--hardware interaction modeling.

Fig.~\ref{fig:ablation} shows that \crest{} achieves the highest mean fidelity in all seven datasets. The clearest gains occur on Quantum Volume, Q-Tetris, and MQT Bench: their means increase from 0.37205 to 0.39891, from 0.17333 to 0.19361, and from 0.16850 to 0.18443, respectively. These correspond to improvements of 7.22\%, 11.70\%, and 9.45\%. The benefit spans randomized interactions, application circuits, and several algorithmic structures, supporting the use of circuit--region relationships beyond global quality summaries.

Aggregate statistics describe circuit scale and regional quality compactly, but they compress where interactions occur and how gate dependencies organize them. Regions with similar summary features can therefore support a particular circuit differently. The graph encoders preserve these relationships, and bidirectional attention relates the circuit's interaction demands to local hardware resources before scoring. The observed gains support this joint structural representation as a useful addition to the statistics-based path, especially where global summaries alone leave the matching between computation and resources unresolved.

\subsection{Generalization Evaluation}
\label{sec:generalization}

\begin{figure}[!t]
\centering
\includegraphics[width=0.98\linewidth]{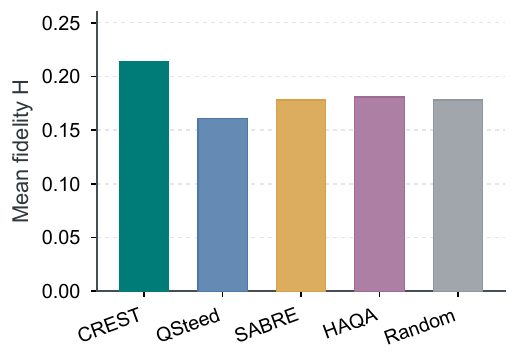}
\caption{Mean fidelity of the five compilation flows on the 96 circuits held out from the TianYan-176 real-device development splits. Each method is evaluated on the same held-out circuits.}
\label{fig:generalization}
\end{figure}

To test whether the learned scoring rule transfers beyond the circuits used to develop it, we reserve 96 circuits from all seven datasets for evaluation only. These circuits take no part in parameter training or model selection. The trained model chooses their regions using the same candidate-construction and routing procedure as in the main evaluation. Thus, the comparison tests selection on unseen circuit instances while holding the compilation conditions fixed.

As shown in Fig.~\ref{fig:generalization}, \crest{} achieves mean fidelity of 0.21347, compared with 0.18100 for HAQA, the strongest baseline on the held-out subset. This corresponds to a 17.94\% improvement, close to the 19.03\% gain on the complete suite. The paired mean difference is 0.03247, with a 95\% bootstrap interval of [0.01274, 0.05317]. \crest{} also outperforms QSteed, SABRE, and Random, and their paired difference intervals likewise remain above zero.

The advantage persists under controlled selection conditions. With a common SABRE backend, \crest{} exceeds QSteed by 0.02386. When both the candidate menu and routing backend are shared, the difference is 0.03189, with a 95\% bootstrap interval of [0.01392, 0.05014]. These matched comparisons support the benefit of learned region scoring on unseen circuits. Together with the comparable improvements on the complete and held-out suites, they indicate that the learned selection rule transfers beyond the circuit instances used for model development.

\subsection{Robust Region Selection across Calibration Updates}
\label{sec:repeatability}

\begin{figure}[!tp]
\centering
\includegraphics[width=0.98\linewidth]{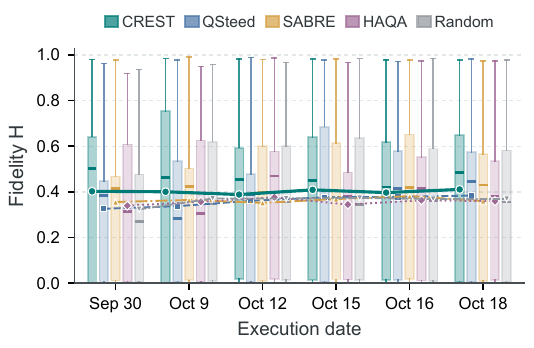}
\caption{Fidelity distributions for QASMBench over 17 matched circuits at six measurement dates. Connected markers show means; hollow points show statistical outliers. Boxes span the first and third quartiles, center lines show medians, and whiskers extend to observations within 1.5 interquartile ranges. All outliers are retained.}
\label{fig:repeatability_qasmbench}
\end{figure}

\begin{figure}[!tp]
\centering
\includegraphics[width=0.98\linewidth]{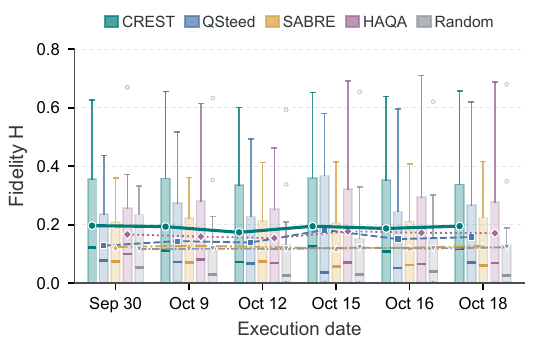}
\caption{Fidelity distributions for Q-Tetris over 12 matched circuits at six measurement dates. Connected markers show means; hollow points show statistical outliers.}
\label{fig:repeatability_q_tetris}
\end{figure}

\begin{figure}[!tp]
\centering
\includegraphics[width=0.98\linewidth]{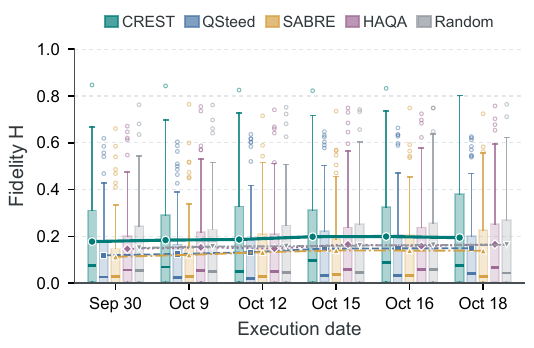}
\caption{Fidelity distributions for MQT Bench over 61 matched circuits at six measurement dates. Connected markers show means; hollow points show statistical outliers.}
\label{fig:repeatability_mqt_bench}
\end{figure}

\begin{figure}[!tp]
\centering
\includegraphics[width=0.98\linewidth]{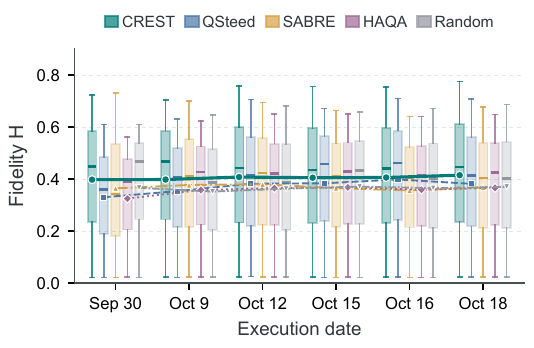}
\caption{Fidelity distributions for Quantum Volume over 3 matched circuits at six measurement dates. Connected markers show means; hollow points show statistical outliers.}
\label{fig:repeatability_quantum_volume}
\end{figure}

\begin{figure}[!tp]
\centering
\includegraphics[width=0.98\linewidth]{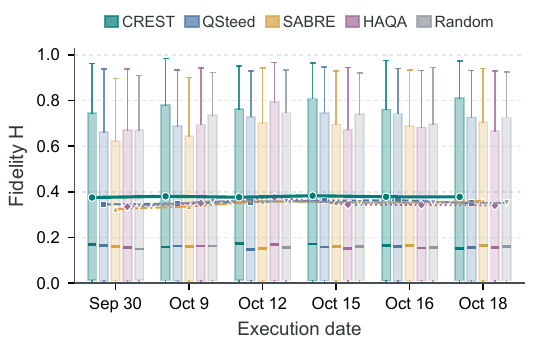}
\caption{Fidelity distributions for HW-efficient VQE over 20 matched circuits at six measurement dates. Connected markers show means; hollow points show statistical outliers.}
\label{fig:repeatability_hw_efficient_vqe}
\end{figure}

\begin{figure}[!tp]
\centering
\includegraphics[width=0.98\linewidth]{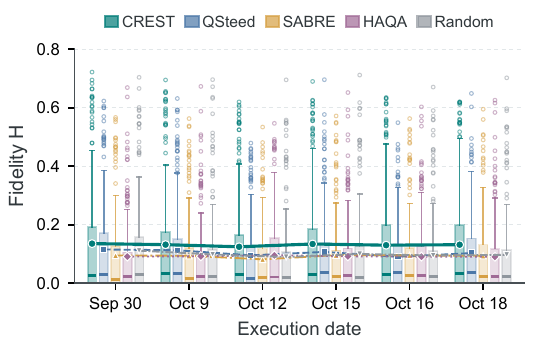}
\caption{Fidelity distributions for Static QEC over 150 matched circuits at six measurement dates. Connected markers show means; hollow points show statistical outliers.}
\label{fig:repeatability_static_qec}
\end{figure}

\begin{figure}[!tp]
\centering
\includegraphics[width=0.98\linewidth]{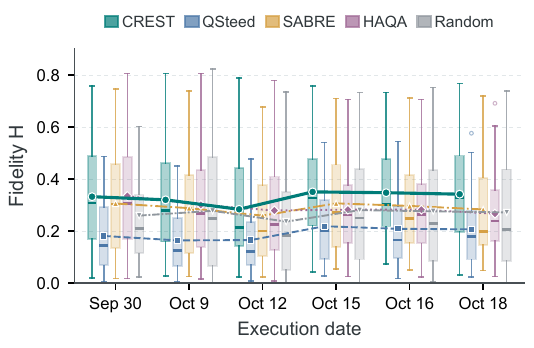}
\caption{Fidelity distributions for Three-regular over 60 matched circuits at six measurement dates. Connected markers show means; hollow points show statistical outliers.}
\label{fig:repeatability_three_regular}
\end{figure}
\begin{table*}[!t]
\centering
\caption{Dataset-Level Fidelity at Six Measurement Dates. Best baseline denotes the highest dataset mean among QSteed, SABRE, HAQA, and Random.}
\label{tab:repeatability}
\footnotesize
\setlength{\tabcolsep}{1.2pt}
\begin{tabular}{@{}lr rr rr rr rr rr rr@{}}
\toprule
 & & \multicolumn{2}{c}{Sep. 30, 2026} & \multicolumn{2}{c}{Oct. 9, 2026} & \multicolumn{2}{c}{Oct. 12, 2026} & \multicolumn{2}{c}{Oct. 15, 2026} & \multicolumn{2}{c}{Oct. 16, 2026} & \multicolumn{2}{c}{Oct. 18, 2026} \\
\cmidrule(lr){3-4}\cmidrule(lr){5-6}\cmidrule(lr){7-8}\cmidrule(lr){9-10}\cmidrule(lr){11-12}\cmidrule(lr){13-14}
Dataset & $N$ & CREST & Baseline & CREST & Baseline & CREST & Baseline & CREST & Baseline & CREST & Baseline & CREST & Baseline \\
\midrule
QASMBench & 17 & \textbf{0.40285} & 0.35597 & \textbf{0.40077} & 0.37116 & \textbf{0.38864} & 0.37668 & \textbf{0.40909} & 0.37640 & \textbf{0.39714} & 0.38146 & \textbf{0.41101} & 0.38418 \\
Q-Tetris & 12 & \textbf{0.19706} & 0.16701 & \textbf{0.19361} & 0.15931 & \textbf{0.17395} & 0.15436 & \textbf{0.19493} & 0.18237 & \textbf{0.18737} & 0.17265 & \textbf{0.19530} & 0.17122 \\
MQT Bench & 61 & \textbf{0.17812} & 0.15100 & \textbf{0.18443} & 0.15695 & \textbf{0.18701} & 0.15811 & \textbf{0.19892} & 0.16591 & \textbf{0.19970} & 0.16352 & \textbf{0.19502} & 0.16626 \\
Quantum Volume & 3 & \textbf{0.39840} & 0.36703 & \textbf{0.39891} & 0.37803 & \textbf{0.40848} & 0.38277 & \textbf{0.40602} & 0.38448 & \textbf{0.40673} & 0.39917 & \textbf{0.41572} & 0.38243 \\
HW-efficient VQE & 20 & \textbf{0.37565} & 0.34513 & \textbf{0.38095} & 0.35930 & \textbf{0.37685} & 0.37647 & \textbf{0.38342} & 0.36155 & \textbf{0.37874} & 0.36397 & \textbf{0.37830} & 0.36012 \\
Static QEC & 150 & \textbf{0.13554} & 0.11546 & \textbf{0.13199} & 0.11285 & \textbf{0.12465} & 0.09553 & \textbf{0.13423} & 0.10717 & \textbf{0.13015} & 0.09532 & \textbf{0.13182} & 0.10311 \\
Three-regular & 60 & 0.33302 & \textbf{0.33563} & \textbf{0.32085} & 0.30155 & \textbf{0.28447} & 0.27982 & \textbf{0.35140} & 0.30730 & \textbf{0.34828} & 0.29745 & \textbf{0.34260} & 0.28390 \\
\bottomrule
\end{tabular}
\end{table*}

To evaluate robustness across calibration updates, we test whether a fixed \crest{} model retains its execution-fidelity advantage when supplied with the hardware information available at each measurement date. We reuse the same model on September 30 and October 9, 12, 15, 16, and 18, 2026, without retraining or changing its parameters. The longitudinal comparison uses the same 323 circuits with results for every method at all six dates, including 95 circuits from the original holdout.

Figs.~\ref{fig:repeatability_qasmbench}--\ref{fig:repeatability_three_regular} show the seven datasets on this matched set. Connected markers give the means; boxes and whiskers show the circuit-level distributions, with every outlier retained. Table~\ref{tab:repeatability} compares \crest{} with the strongest dataset mean among the baselines at each date.

\crest{} has the highest overall mean at all six dates, with mean fidelities ranging from 0.20009 to 0.22147. Its advantage over the strongest overall baseline remains positive throughout, ranging from 13.34\% to 24.06\%. It also leads in 41 of the 42 dataset--date comparisons in Table~\ref{tab:repeatability}. Absolute fidelity and individual margins vary with the hardware execution, but the learned selection rule retains a broad advantage across both dates and workloads. On October 18, the common-set mean is 0.21786, compared with 0.17605 for Random; their paired difference is 0.04181, with a 95\% bootstrap interval of [0.03007, 0.05336].

The complete October 18 suite contains all 324 circuits and reaches mean fidelity 0.21848, exceeding the strongest baseline at 0.17686, by 23.53\%. On all 96 original held-out circuits, \crest{} reaches 0.21585 versus 0.17517 for HAQA, an improvement of 23.22\%. The longitudinal 95-circuit holdout gives 0.21371 versus 0.17224 on that date, a 24.07\% advantage. The favorable temporal comparison therefore also extends to circuits excluded from model development.

The sustained relative fidelity advantage provides evidence of robust region selection across these calibration updates. Current qubit and coupler attributes change the resource representations, while the learned circuit--hardware interaction model is reused. These results support retaining useful selection knowledge as hardware conditions evolve, extending the framework's execution-quality benefit beyond a single measurement date.

\subsection{Computational Overhead}
\label{sec:overhead}

\begin{table}[!t]
\centering
\caption{Per-Circuit Online Processing Time on the 324-Circuit Suite, in Seconds}
\label{tab:timing}
\small
\begin{tabular}{@{}lrrr@{}}
\toprule
Method & Mean & Median & 95th percentile \\
\midrule
CREST & 0.168 & 0.147 & 0.302 \\
QSteed & 0.187 & 0.108 & 0.577 \\
SABRE & 0.039 & 0.023 & 0.118 \\
HAQA & 0.033 & 0.019 & 0.118 \\
Random & 0.033 & 0.019 & 0.099 \\
\bottomrule
\end{tabular}
\end{table}

\crest{} separates calibration-dependent offline preparation from per-circuit online processing. The offline stage constructs candidate regions and caches their hardware features and circuit-independent GNN representations. This preparation has a modest cost: candidate construction and hardware-feature preparation average only 1.295 seconds per width for circuits with 4--20 qubits, while precomputing the GNN representations takes an additional 4.108 seconds for the full candidate catalog. These resources can be prepared before circuits arrive and reused under the same calibration.

The online stage includes circuit parsing and feature extraction, GNN-based region scoring and selection, and compilation through layout, routing, and native-gate conversion to produce the output circuit and physical mapping. With Python and ONNX Runtime on CPU, \crest{} completes this entire stage in 0.168 seconds per circuit on average over the 324-circuit suite. Table~\ref{tab:timing} summarizes the online processing times of all five methods.


\section{Discussion and Conclusion}
\label{sec:conclusion}

This work presents \crest{}, a hardware-aware, GNN-based framework for high-fidelity superconducting quantum compilation. It integrates multiscale candidate-region construction, learned circuit--region scoring, and in-region compilation. Hierarchical circuit representations and calibrated region graphs preserve computational demands and resource characteristics, while bidirectional cross-graph attention relates them for fidelity-oriented selection. The trained evaluator is reused across calibration updates by refreshing the hardware attributes supplied to scoring.

Evaluations on 324 circuits from seven datasets show that \crest{} improves mean measurement-distribution fidelity by 19.03\% over the strongest overall baseline. With identical candidates and routing, its improvement over QSteed reaches 18.74\%, supporting the effectiveness of learned region scoring. The statistics-only ablation further supports the benefit of structured circuit--hardware modeling, while the held-out evaluation demonstrates transfer to unseen circuit instances. Moreover, without retraining, the same model maintains the highest overall mean fidelity across six measurement dates, supporting robust region selection as hardware calibration evolves.

By linking fidelity-oriented region choices with reuse across calibration updates, \crest{} provides a learning-based approach to compiling diverse quantum workloads under evolving hardware conditions. Future research can extend this framework to region selection for dynamic circuits, transfer across processor architectures, and joint optimization of region selection and routing.

\bibliographystyle{IEEEtran}
\bibliography{crestref_etal}

\end{document}